\documentclass[preprint,12pt]{elsarticle}

\usepackage{amssymb}
\usepackage{amsmath}

\usepackage{booktabs}
\usepackage{hyperref}
\journal{Nuclear Instruments and Methods in Physics Research Section A}

\begin{document}

\begin{frontmatter}



\title{Evaluation of PID Performance at CEPC and Optimization \\ with Combined dN/dx and Time-of-Flight Data}


\author[label1,label2,label3,label4,label5]{Dian Yu\fnref{fn1}}
\ead[url]{https://orcid.org/0000-0002-6789-020X}

\author[label6]{Houqian Ding\fnref{fn1}}
\fntext[fn1]{These authors contributed equally to this work.}

\author[label7]{Yongfeng Zhu}

\author[label1,label2,label3,label4]{Kun Liu}
\ead[url]{https://orcid.org/0000-0001-5807-0501}

\author[label6]{Ming Qi}
\ead[url]{https://orcid.org/0000-0002-9221-0683}

\author[label7]{Yunyun Fan\corref{cor1}}
\ead{fanyy@ihep.ac.cn}
\ead[url]{https://orcid.org/0000-0001-7868-3858}
\cortext[cor1]{Corresponding author}

\affiliation[label1]{organization={Tsung-Dao Lee Institute, Shanghai Jiao Tong University},
            city={Shanghai},
            postcode={201210}, 
            country={China}}
\affiliation[label2]{organization={Institute of Nuclear and Particle Physics, School of Physics and Astronomy, Shanghai Jiao Tong University},
            city={Shanghai},
            postcode={200240}, 
            country={China}}
\affiliation[label3]{organization={Key Laboratory for Particle Astrophysics and Cosmology (MOE), Shanghai Key Laboratory for Particle Physics and Cosmology (SKLPPC)},
            city={Shanghai},
            postcode={200240}, 
            country={China}}
\affiliation[label4]{organization={State Key Laboratory of Dark Matter Physics},
            city={Shanghai},
            postcode={200240}, 
            country={China}}
\affiliation[label5]{organization={Universit\'e Paris Cit\'e, CNRS, Astroparticule et Cosmologie},
            city={Paris},
            postcode={75015}, 
            country={France}}
\affiliation[label6]{organization={Department of Physics, Nanjing University},
            city={Nanjing},
            postcode={210093}, 
            country={China}}
\affiliation[label7]{organization={Institute of High Energy Physics, CAS},
            city={Beijing},
            postcode={100049}, 
            country={China}}
\begin{abstract}
Charged-hadron identification (PID) is a critical requirement for the physics program of the Circular Electron–Positron Collider (CEPC). The baseline detector relies on ionization measurements from a time projection chamber (TPC), which provides strong PID capability at low momenta but becomes less effective at higher momenta. In this work, we investigate an extended PID strategy that combines dN/dx information from the TPC with time-of-flight (ToF) measurements from a silicon-based outer tracker (OTK) and a timing-upgraded inner tracker (ITK) equipped with AC-LGAD sensors.

A unified discriminant is constructed to exploit the complementary sensitivity of ionization and timing observables. The performance is evaluated using simulated $Z \to q\bar{q}$ events, focusing on kaon identification in the presence of dominant pion backgrounds. The combined configuration significantly improves both efficiency and purity over a broad kinematic range, extending PID capability to both sub-GeV and multi-GeV regions.

These results highlight the impact of precision timing on tracking detectors and demonstrate a viable path to enhanced PID performance for future lepton colliders.
\end{abstract}


\begin{highlights}
\item A combined charged-hadron PID framework based on TPC dN/dx and silicon ToF is developed for CEPC.
\item The complementarity between ionization and timing information is quantified over a wide momentum range.
\item An upgraded ITK timing layer substantially improves PID performance below 1~GeV/$c$.
\item The combined ITK+TPC+OTK configuration achieves 97.1\% kaon efficiency and 85.6\% purity.
\end{highlights}

\begin{keyword}


CEPC \sep particle identification \sep dN/dx \sep time-of-flight \sep TPC \sep silicon timing detector
\end{keyword}

\end{frontmatter}



\section{Introduction}
\label{sec:introduction}

The Circular Electron--Positron Collider (CEPC) is a proposed next-generation high-luminosity $e^+e^-$ collider designed for precision Higgs and electroweak measurements~\cite{thecepcstudygroup2025cepcreftdr}. With a circumference of 100~km and two interaction points, the CEPC is planned to operate at several center-of-mass energies corresponding to distinct physics goals. These include 240~GeV for Higgs boson production, 160~GeV for a $W^{+}W^{-}$ threshold scan, 91~GeV for $Z$-pole operation, and 360~GeV for top-quark pair production. The baseline operational parameters and expected boson yields are summarized in Table~\ref{tab:cepc-scheme} and detailed in Ref.~\cite{thecepcstudygroup2025cepcreftdr}. The CEPC infrastructure is also designed to accommodate a future upgrade to a proton--proton collider with a center-of-mass energy of up to 75~TeV, enabling direct searches for physics beyond the Standard Model.

\begin{table}[t]
    \centering
    \caption{The CEPC baseline operation scheme, including center-of-mass energies ($\sqrt{s}$), power levels, instantaneous luminosity ($\mathcal{L}$), total integrated luminosity ($\int\mathcal{L}$), and event yields under different operation modes. (*) The maximum instantaneous luminosity achieved during $Z$-pole operation depends on the detector solenoid magnetic field. The value reported here assumes a 3~T solenoid. For a 2~T magnet, the luminosity will be 78\% higher. ($\dagger$) The cross section of $W^{+}W^{-}$ at threshold is about 8~pb. An additional $1.5 \times10^8 W^{+}W^{-}$ events will be produced during the Higgs factory operation~\cite{thecepcstudygroup2025cepcreftdr}.}
    \label{tab:cepc-scheme}
    
    \begin{tabular}{lccc}
        \toprule
        Operation mode & $H$ & $Z$ & $W^{+}W^{-}$ \\
        \midrule
        $\sqrt{s}$ [GeV] & 240 & 91 & $155$--$170$ \\
        SR power [MW] & 30 & 12.1 & 30 \\
        $\mathcal{L}$ [$\times10^{34}\,\mathrm{cm}^{-2}\mathrm{s}^{-1}$] & 5 & 26(*) & 16 \\
        $\int \mathcal{L}$/year [ab$^{-1}$, 2 IPs] & 0.65 & 3.2 & 1.2 \\
        Years & 15 & 4 & 1 \\
        Total $\int \mathcal{L}\,dt$ [ab$^{-1}$, 2 IPs] & 10 & 13 & 1.2 \\
        Event yields [2 IPs] & $2\times10^6$ & $5.6\times10^{11}$ & $1.0\times10^7$($\dagger$) \\
        \bottomrule
    \end{tabular}
\end{table}

A primary objective of the CEPC is the precise measurement of Higgs and electroweak observables, which places stringent requirements on detector performance, including flavor tagging, lepton identification, and charged-hadron particle identification (PID). The large sample of hadronic $Z \to q\bar{q}$ decays produced during $Z$-pole operation provides an ideal environment for detector calibration and for detailed performance studies. In particular, it offers a unique opportunity to develop and validate PID strategies relevant to both precision measurements and future flavor-physics analyses. In this work, we focus on the identification of charged hadrons in such events.

The baseline CEPC detector design employs a time projection chamber (TPC) as its primary tracking detector, providing excellent spatial resolution together with ionization information that can be exploited for charged-hadron PID. In this work, we focus on the dN/dx approach, in which the number of primary ionization clusters per unit length is used as the relevant observable. Compared with conventional charge-based dE/dx measurements, dN/dx is expected to offer improved separation performance in gaseous detectors.

To further enhance the PID capability, particularly for tracks beyond the TPC volume, a silicon-based outer tracker (OTK) is included in the detector layout. The OTK provides precise time-of-flight (ToF) measurements and serves as a complementary subsystem that improves particle separation in the intermediate- and high-momentum regions. Although the OTK is part of the baseline geometry, its potential contribution to PID performance has not yet been systematically quantified.

In addition, we consider an upgraded configuration of the inner silicon tracker (ITK), in which the outermost layer is equipped with AC-LGAD sensors providing precise timing information. This extension enables time-of-flight (ToF) measurements for low-momentum tracks that do not reach the OTK, thereby extending the PID coverage into the sub-GeV region.

In this paper, we present a detailed evaluation of charged-hadron PID performance at the CEPC by combining dN/dx measurements from the TPC with ToF information from both the OTK and the upgraded ITK. A unified PID discriminant is constructed on a track-by-track basis, and the identification regions are optimized in bins of momentum and polar angle by maximizing the product of kaon identification efficiency and purity. The impact of pion contamination, which constitutes the dominant background in hadronic final states, is quantitatively evaluated.

The rest of the paper is organized as follows. Section~\ref{sec:refdetector} introduces the detector geometry, software framework, and simulation samples. Section~\ref{sec:separationevaluation} presents the expected separation power. Section~\ref{sec:pidperformance} describes the PID strategy and optimization procedure. The resulting performance is discussed in Section~\ref{sec:results}, followed by the conclusions in Section~\ref{sec:conclusion}.

\section{Detector configuration, software framework, and simulated samples}
\label{sec:refdetector}

\subsection{Detector overview}
\label{sec:detectoroverview}

\begin{figure}[t]
    \centering
    \includegraphics[width=\linewidth]{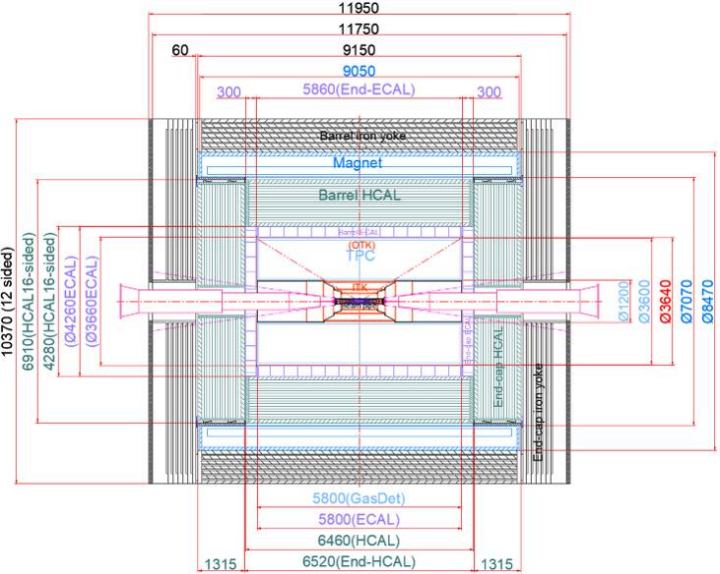}
    \caption{The geometry of the baseline CEPC design~\cite{thecepcstudygroup2025cepcreftdr}.}
    \label{fig:geometry}
\end{figure}

The layout of the baseline CEPC detector design is shown in Fig.~\ref{fig:geometry}~\cite{thecepcstudygroup2025cepcreftdr}. It is designed following the particle flow principle~\cite{ruan2014arbornewapproachparticle}, which emphasizes the separation of final state particles and measures each final state particle in the most suitable sub-detector. From the innermost to the outermost layer, the CEPC detector consists of a three-layer double-sided silicon detector used for vertex reconstruction, a three-layer silicon-based inner tracker, a TPC detector, a silicon outer tracker layer, a silicon-tungsten sampling electromagnetic calorimeter, a steel-glass hadronic calorimeter based on resistive plate chambers, a 3~T superconducting solenoid and a flux return yoke embedded with a muon detector.

\subsection{The TPC detector}
\label{sec:tpc}

\begin{figure}[t]
    \centering
    \includegraphics[width=\linewidth]{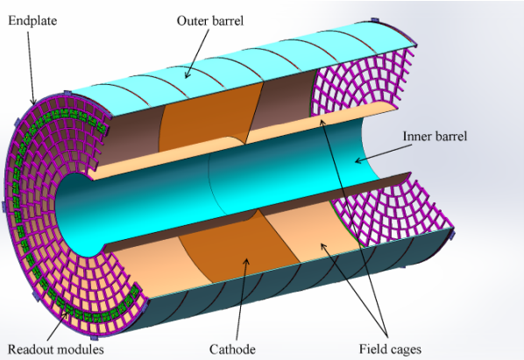}
    \caption{ The TPC is a cylindrical, gas-filled tracking device whose axis coincides with the nominal beam direction. The barrel encloses a field cage assembly that produces a uniform electric drift field along the z-axis. Two readout endplates mechanically support the barrels, while the central cathode plane is positioned at the barrel’s mid-plane.~\cite{thecepcstudygroup2025cepcreftdr}.}
    \label{fig:tpc}
\end{figure}

Fig.~\ref{fig:tpc} shows a sketch of the TPC detector~\cite{thecepcstudygroup2025cepcreftdr}. The CEPC TPC is a gaseous cylindrical tracking detector based on a field-cage structure. The inner and outer radii of the TPC are 0.6~m and 1.8~m, respectively, and the full length of the TPC detector is 5.8~m. 
A working gas mixture characterized by low transverse diffusion and high drift velocity, identical to the T2K gas mixture $(\mathrm{Ar}/\mathrm{CF}_4/\mathrm{iC}_4\mathrm{H}_{10}=95/3/2)$, is adopted.
The central cathode plane, shown in dark yellow in Fig.~\ref{fig:tpc}, is held at an electric potential of $-63~\mathrm{kV}$, while the end plates, shown in purple, are kept at ground potential, thereby establishing a highly homogeneous electric field of 230~V/cm that drives the ionized electrons toward the readout. Under this drift electric field, the maximum drift time of the ionized electrons around the cathode in the T2K gas is about 34~$\mu$s.

In addition to the electric field, a 3~T solenoidal magnetic field is applied to suppress transverse diffusion and improve the position resolution.

\begin{figure}[t]
    \centering
    \includegraphics[width=\linewidth]{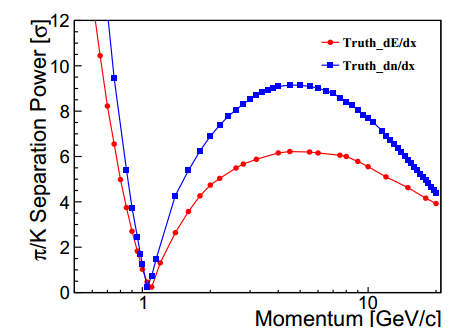}
    \caption{$K$--$p/\bar{p}$ separation power in a gas detector using the dN/dx and dE/dx methods as a function of momentum, obtained from MC truth values in Garfield++~\cite{gaseousdetectorPID}.}
    \label{fig:dndxvsdedx}
\end{figure}

Fig.~\ref{fig:dndxvsdedx} compares the PID performance achievable with the dN/dx and dE/dx methods. In the dN/dx method, the measured value of dN/dx represents the number of initial ionization clusters per unit distance for different particles, while in the dE/dx method, it denotes charge per unit distance. Simulation studies demonstrate that the dN/dx method provides superior PID performance compared to the dE/dx method.

\begin{figure}[t]
    \includegraphics[width=\linewidth]{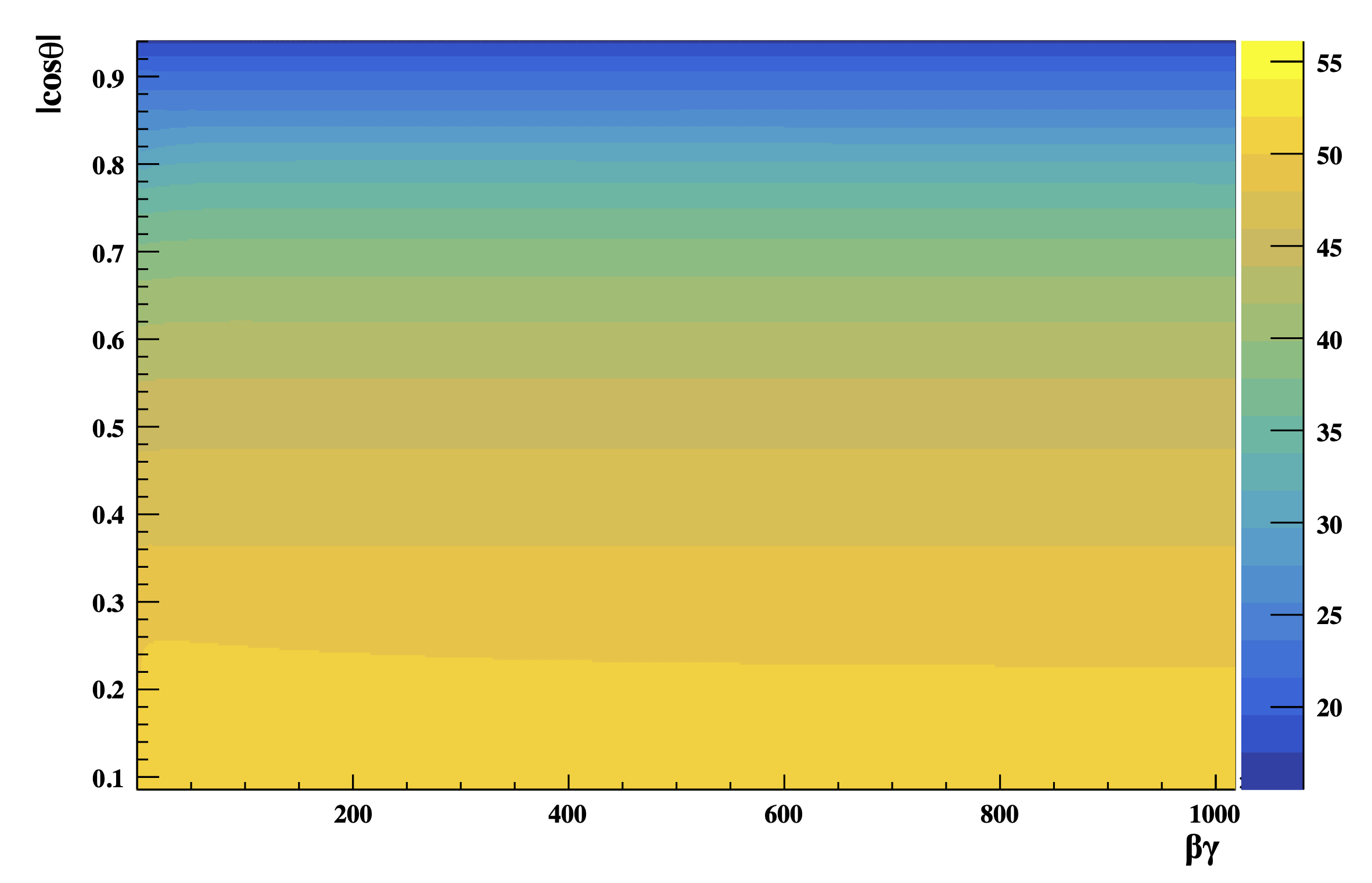}
    {\centering (a) \par}
    \vspace{1em}
    \includegraphics[width=\linewidth]{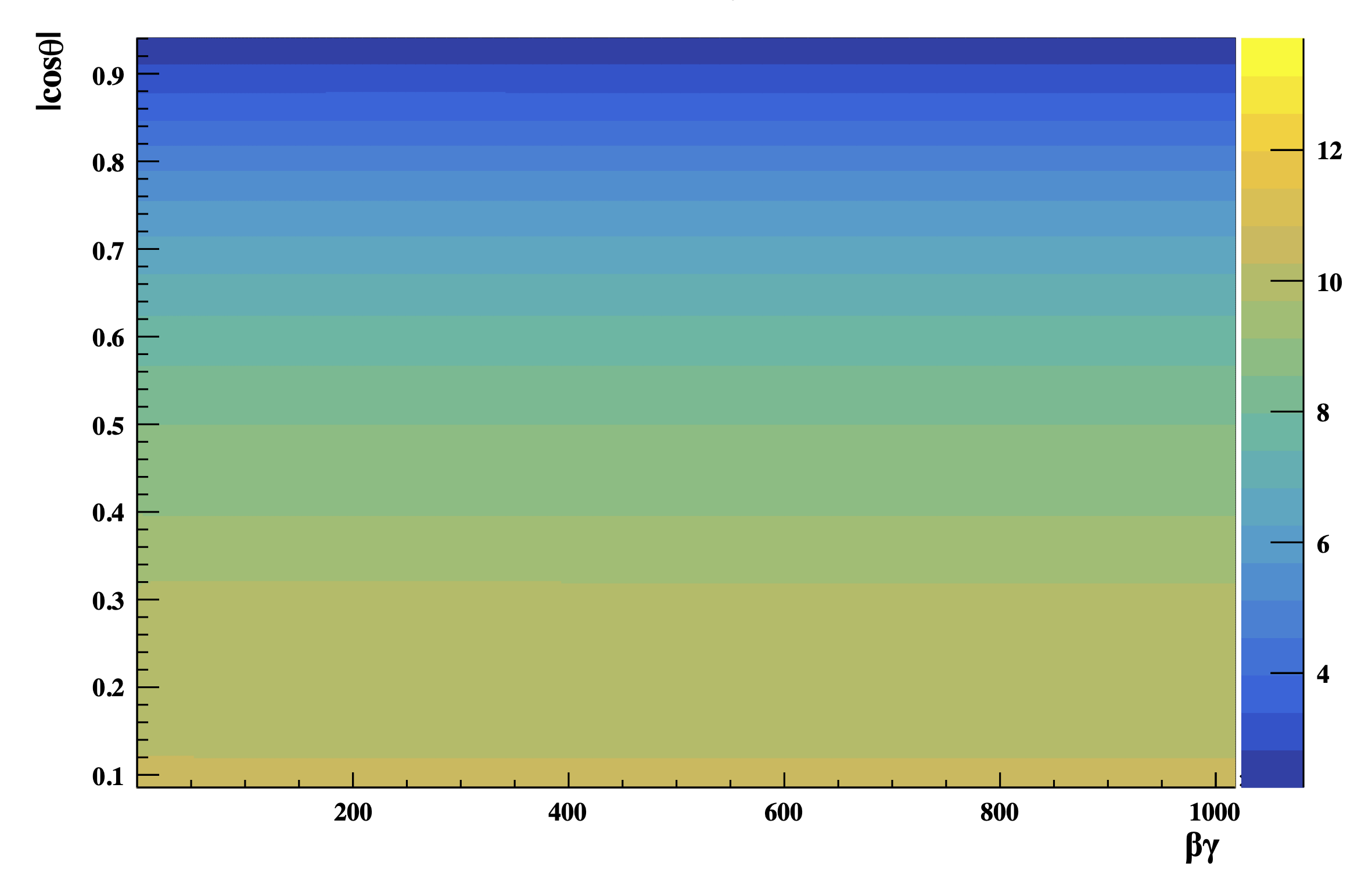}
    {\centering (b) \par}
    \caption{\label{fig:tpc_value}%
    TPC readout dN/dx mean and sigma as a function of particle angle and velocity.\\
    (a): Readout dN/dx mean.\\
    (b): Readout dN/dx sigma~\cite{Qi:2023PixelTPC}.}
\end{figure}

Fig.~\ref{fig:tpc_value} shows two look-up tables (LUT) of the TPC readout dN/dx mean and sigma for incident particles with given angle and velocity~\cite{Qi:2023PixelTPC}. The LUTs are generated based on particle gun simulation results using Garfield++~\cite{Garfieldpp}. For each gun configuration, the detector response is fitted with a Gaussian function. The resulting mean and standard deviation are taken as the ideal readout dN/dx mean and sigma of the corresponding particle. The particle angle and velocity are used as inputs to query the corresponding dN/dx mean and sigma from the LUTs. The final output is obtained by applying a Gaussian smearing based on these values.

\subsection{Timing silicon detectors}
To enhance the particle identification (PID) capability of the baseline CEPC detector, this study introduces dedicated timing layers at both the inner and outer regions of the tracking system. Specifically, two silicon-based time-of-flight (ToF) detectors are considered: the Inner Tracker (ITK) and the Outer Tracker (OTK), both designed using low-gain avalanche detector (AC-LGAD) technology.

The OTK is implemented as a strip-type AC-LGAD detector placed just outside the TPC and is designed to provide precise timing measurements for tracks in the intermediate- and high-momentum regions. Since the main function of the OTK is to provide ToF information, fine spatial resolution along the $z$-axis (beam direction) is not required. The chosen design achieves a spatial resolution of 10~$\mu\mathrm{m}$ and a timing resolution of 30~ps.

Complementarily, the ITK is proposed to be upgraded with a pixelated AC-LGAD layer at its outermost radius. This inner timing layer addresses the lack of ToF coverage for low-momentum particles, which typically do not reach the outer detector. The pixel configuration ensures sufficient spatial granularity for precise tracking and allows ToF measurements close to the interaction point.

In this study, both the ITK and OTK are treated as integral innovations designed to extend PID capability across the full momentum range. Unless otherwise specified, ITK and OTK refer to these timing-enhanced configurations.

\subsection{Software}
A dedicated software framework, CEPCSW, has been developed to support the full simulation, reconstruction, and analysis workflow of the CEPC experiment. CEPCSW is built upon Gaudi~\cite{gaudi} and DD4hep~\cite{dd4hep}, providing a modular and extensible environment for high-energy physics applications. The detector geometry and conditions data are described using DD4hep, while the full simulation is implemented through Geant4~\cite{geant4} via the k4SimGeant4 interface. The framework uses EDM4hep as the event data model and supports digitization, tracking, calorimetry, and particle flow reconstruction. CEPCSW is hosted and actively maintained on the IHEP GitLab~\cite{cepcsw}, and it serves as the basis for physics performance studies and detector R\&D at CEPC. Version 25.3.6 is used in this analysis.

\subsection{Samples}
Using CEPC baseline detector geometry and CEPCSW, we simulated 8.83$\times$10$^{\text{6}}$ hadronic $Z$-pole events, which is equivalent to 1.26$\times$10$^{\text{7}}$ inclusive $Z$-pole events containing the non-hadronic events. These events encompass all the different quark flavors, as predicted by the SM, with details presented in Table~\ref{tab:Zhard}.

\begin{table}[t]
    \centering
    \caption{The branching ratio and the number of simulated events of the $Z$-pole samples.
    }
    \label{tab:Zhard}
    \begin{tabular}{ccc}
        \toprule
        Process & Branching ratio & Total events \\
        \midrule
        $Z\rightarrow u\bar{u}$ & 11.17\% &  \\
        $Z\rightarrow d\bar{d}$ & 15.84\% &  \\
        $Z\rightarrow s\bar{s}$ & 15.84\% & 1.26$\times10^7$ \\
        $Z\rightarrow c\bar{c}$ & 12.03\% &  \\
        $Z\rightarrow b\bar{b}$ & 15.12\% &  \\
        \bottomrule
    \end{tabular}
\end{table}

\section{Separation evaluation}
\label{sec:separationevaluation}

\subsection{Separation power definition}
\label{separationpowerdefinition}

This section presents the expected ToF performance of the ITK and OTK together with the dN/dx performance of the TPC, and evaluates the corresponding separation power for $K^{\pm}$, $\pi^{\pm}$, and $p/\bar{p}$. The PID capability is quantified using separation power, defined as

\begin{equation}
S_{\text{AB}}=\frac{|\mu_\text{A} - \mu_\text{B}|}{(\sigma_\text{A}+\sigma_\text{B})/2}
\end{equation}

where $\mu_{\text{A(B)}}$ denotes the mean measured value of the observable (ToF or dN/dx) of particle A(B) and $\sigma_{\text{A(B)}}$ represents the corresponding measurement resolution. Separation powers can be combined as a quadrature sum, $S_\text{combine}=S_\text{TPC}\oplus S_\text{TOF}$.

\subsection{Separation power of ToF}
\label{separationpoweroftof}

The ToF of a charged particle is determined by its momentum and by the flight distance along its trajectory. In the presence of a magnetic field, the trajectory is helical and may intersect the timing detector either in the barrel or in the end-cap region. The radius of the helix is given by

\begin{equation*}
r=\frac{1000\,P_{\mathrm{t}}}{0.3\,B\,q}
\end{equation*}

where $P_{\mathrm{t}}$, $q$, and $B$ denote the transverse momentum, charge, and magnetic-field strength, respectively.

Let $R_{\mathrm{barrel}}$ denote the inner radius of the barrel timing layer, $R_{\mathrm{end\text{-}cap}}$ the inner radius of the end-cap timing layer, and $L$ the distance between the interaction point and the end-cap layer along the beam direction. The transverse and longitudinal components of the particle velocity are denoted by $v_{\mathrm{t}}$ and $v_{\mathrm{l}}$, respectively. According to the geometry of the helical trajectory, three cases are considered:

\begin{enumerate}
    \item If $r < R_{\mathrm{barrel}}$, the particle reaches the end-cap without intersecting the barrel, and the flight time is
    \begin{equation*}
    \mathrm{ToF} = \frac{L}{v_{\mathrm{l}}}.
    \end{equation*}

    \item If $r > R_{\mathrm{barrel}}$ and the particle reaches the end-cap before intersecting the barrel, the flight time is
    \begin{equation*}
    \mathrm{ToF} = \frac{r\,2\arcsin\!\left(R_{\mathrm{barrel}}/2r\right)}{v_{\mathrm{t}}}.
    \end{equation*}

    \item If $r > R_{\mathrm{barrel}}$ and the particle intersects the barrel before reaching the end-cap, the flight time is
    \begin{equation*}
    \mathrm{ToF} = \frac{L}{v_{\mathrm{l}}}.
    \end{equation*}
\end{enumerate}

\begin{figure}[t]
    \includegraphics[width=\linewidth]{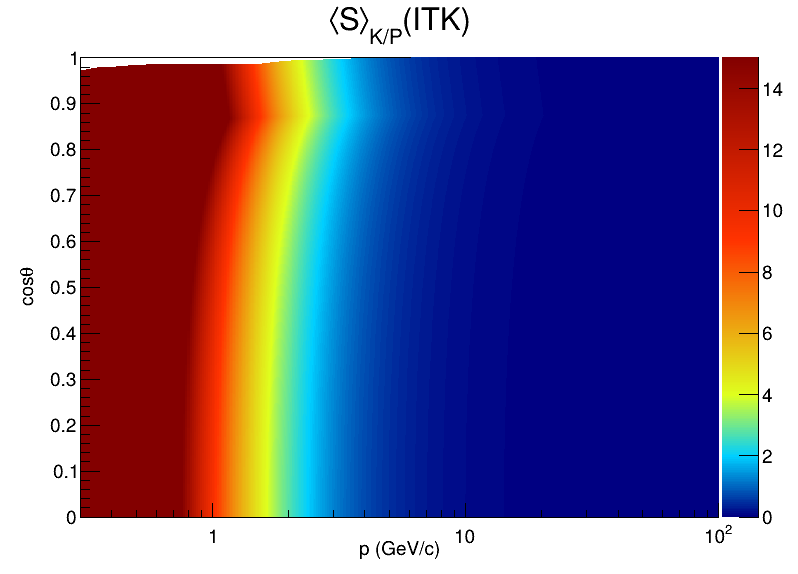}
    {\centering (a) \par}
    \vspace{1em}
    \includegraphics[width=\linewidth]{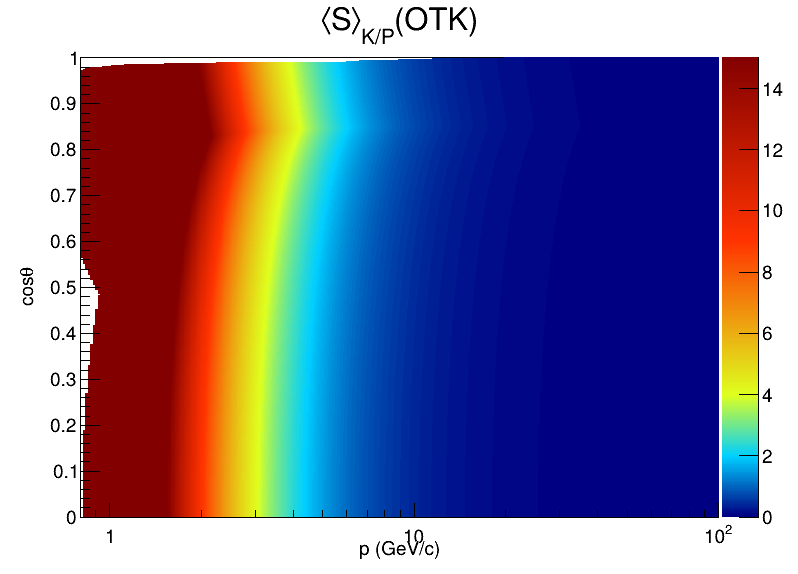}
    {\centering (b) \par}
    \vspace{1em}
    \caption{$K$--$p/\bar{p}$ separation power of the timing detectors as a function of $p$ and $\cos\theta$.\\
    (a) $K$--$p/\bar{p}$ separation power of the ITK.\\
    (b) $K$--$p/\bar{p}$ separation power of the OTK.}
    \label{fig:separationpoweritkotk}
\end{figure}

In the baseline CEPC detector, the OTK serves as the outer timing detector, while the upgraded ITK considered in this work provides timing information at smaller radius. The $K$--$p/\bar{p}$ separation power of the ITK and OTK as a function of momentum and polar angle is shown in Fig.~\ref{fig:separationpoweritkotk}. Regions with separation power greater than 4 are shown in warm colors, those below 2 are shown in cold colors, and the intermediate region from 2 to 4 is shown in transition colors. The OTK achieves $K$--$p/\bar{p}$ separation power above 4 in the momentum range of approximately 1--3~GeV/$c$, while the ITK reaches the same level in the range of approximately 0.3--1.5~GeV/$c$.

\subsection{Separation power of the TPC}
\label{sec:separationpoweroftpc}

\begin{figure}[t]
    \includegraphics[width=\linewidth]{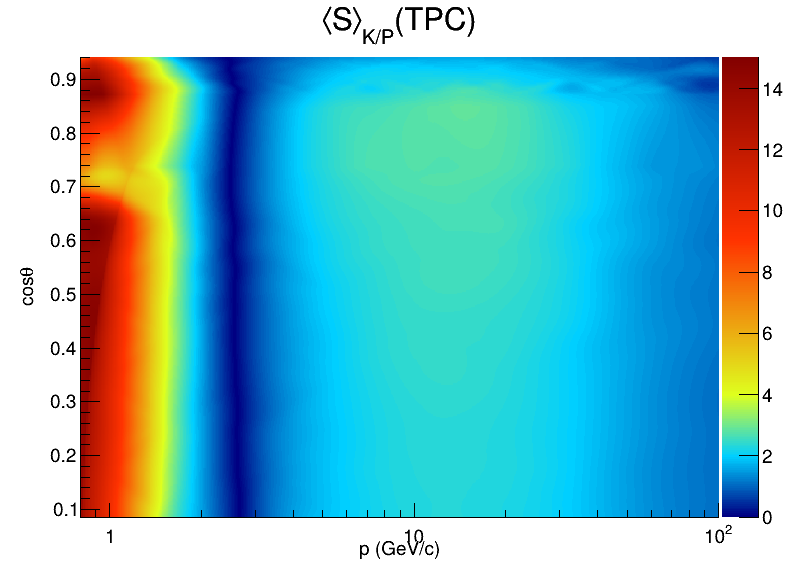}
    {\centering (a) \par}
    \vspace{1em}
    \includegraphics[width=\linewidth]{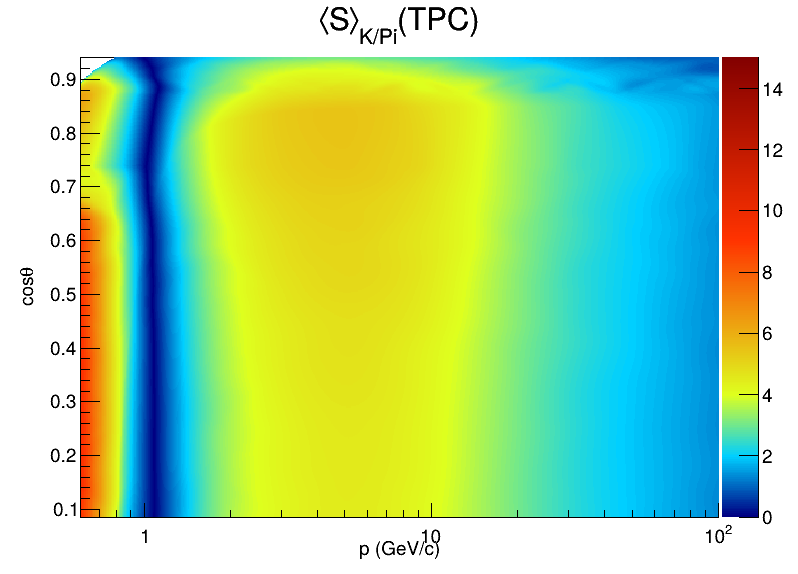}
    {\centering (b) \par}
    \vspace{1em}
    \caption{Separation power of the TPC as a function of $p$ and $\cos\theta$.\\
    (a) $K$--$p/\bar{p}$ separation power of the TPC.\\
    (b) $K$--$\pi$ separation power of the TPC.
    }
    \label{fig:separationpowertpc}
\end{figure}

The separation power of dN/dx is calculated using the definition given in Sec.~\ref{separationpowerdefinition}. The mean dN/dx value is obtained directly from Fig.~\ref{fig:tpc_value}, while the corresponding resolution is derived as

\begin{equation*}
\sigma_{\mathrm{dN/dx}}=\frac{\sigma_{\mathrm{readout}}}{\sqrt{l}}
\end{equation*}

where $l$ is the trajectory length inside the TPC. The calculation of this length is similar to that used for the ToF, except that the starting point is taken to be the intersection between the helix and the inner wall of the TPC. If the point of maximum radial extent lies inside the TPC volume, it is taken as the end point of the trajectory. The resulting separation power of the TPC as a function of momentum and polar angle is shown in Fig.~\ref{fig:separationpowertpc}. The color scheme is the same as in Fig.~\ref{fig:separationpoweritkotk}.

\subsection{Combined separation power}
\label{sec:separationpowercombine}

\begin{figure}[t]
    \includegraphics[width=\linewidth]{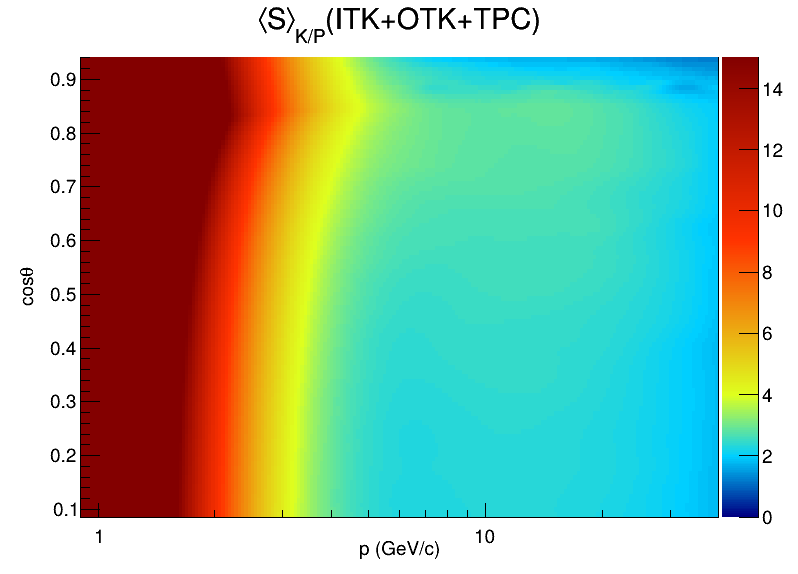}
    {\centering (a) \par}
    \vspace{1em}
    \includegraphics[width=\linewidth]{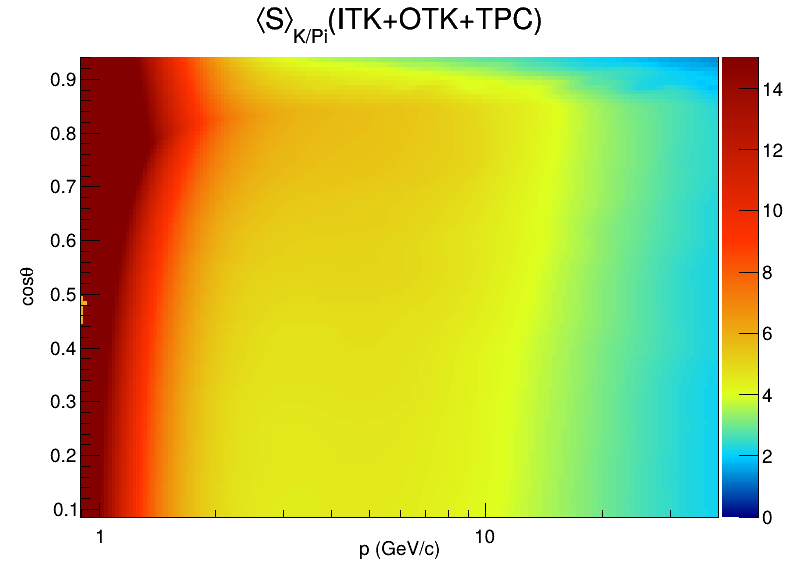}
    {\centering (b) \par}
    \vspace{1em}
    \caption{
    Combined separation power of the ITK, OTK, and TPC as a function of $p$ and $\cos\theta$.\\
    (a) $K$--$p/\bar{p}$ separation power.\\
    (b) $K$--$\pi$ separation power.}
    \label{fig:separationpowercom}
\end{figure}

The above studies evaluate the separation power of $\pi^{\pm}$, $K^{\pm}$, and $p/\bar{p}$ using timing information and TPC information separately. The timing information enhances the TPC separation power around 1~GeV/$c$ for $K$--$p/\bar{p}$ separation and around 2.5~GeV/$c$ for $K$--$\pi$ separation. In particular, the upgraded ITK provides additional separation power in the momentum range from 0.3 to 1.0~GeV/$c$. After combining the TPC and timing information, the resulting separation power as a function of momentum and polar angle for $K$--$p/\bar{p}$ and $K$--$\pi$ is shown in Fig.~\ref{fig:separationpowercom}. The color scheme is the same as in Fig.~\ref{fig:separationpoweritkotk}. The simulation results indicate that the timing detector is complementary to the gaseous detector and significantly improves the particle-separation capability.

\section{PID performance}
\label{sec:pidperformance}

\subsection{PID strategy}
\label{sec:pidstrategy}

The separation-power study in Sec.~\ref{sec:separationevaluation} serves only as an expectation-level illustration. The actual PID assignment in this work is performed event by event using the combined residual $\chi_{\rm PID}$. For each charged track, we consider multiple particle hypotheses ($\pi^{\pm}$, $K^{\pm}$, $p/\bar{p}$). Under each hypothesis, the expected ToF or dN/dx value is computed based on the track momentum and path length. A $\chi$ value is then calculated as

\begin{equation}
\chi_{O}=\frac{O_{\text{meas}}-O_{\text{exp}}}{\sigma_{O}}
\end{equation}

where $O_{\text{meas}}$ is the measured observable, while $O_{\text{exp}}$ denotes the expected value under the given particle hypothesis, $\sigma_{O}$ is the resolution corresponding to the observable. After combining the $\chi$ obtained by each detector, the particle type is assigned according to the region it sits in. The regions are described in the next subsection.

In this article, the combined PID discriminant is given by

\begin{equation}
\chi_{\text{PID}} = \pm\sqrt{\chi^2_{\text{ITKToF}} + \chi^2_{\text{TPC\,dN/dx}} + \chi^2_{\text{OTKToF}}}
\end{equation}

where $\chi_{\text{ITKToF}}$ and $\chi_{\text{OTKToF}}$ are computed using ToF measurements from the ITK and OTK detectors, respectively, and $\chi_{\text{TPC dN/dx}}$ is computed using dN/dx measurements from the TPC. The sign of $\chi_{\mathrm{PID}}$ is determined from the signed sum of the individual residuals. This definition corresponds to a Gaussian likelihood approximation assuming independent detector responses.

The performance of PID is evaluated by efficiency $\epsilon$ and purity $p$, taking $K$ for an example, they are defined by 

\begin{equation}
\epsilon_K=\frac{N_{K\rightarrow K}}{N_K}
\end{equation}

\begin{equation}
p_K=\frac{N_{K\rightarrow K}}{\sum N_{\rightarrow K}}
\end{equation}

where $N_K$, $N_{K\rightarrow K}$, and $\sum N_{\rightarrow K}$ denote the number of true kaons, the number of true kaons identified as kaons, and the total number of tracks identified as kaons, respectively.  It is noted that $\sum N_{\rightarrow K} = N_{K\rightarrow K} + N_{\pi\rightarrow K} + N_{p\rightarrow K}$ in this article.

\subsection{Region selection}

\begin{figure}[t]
    \centering
    \includegraphics[width=\linewidth]{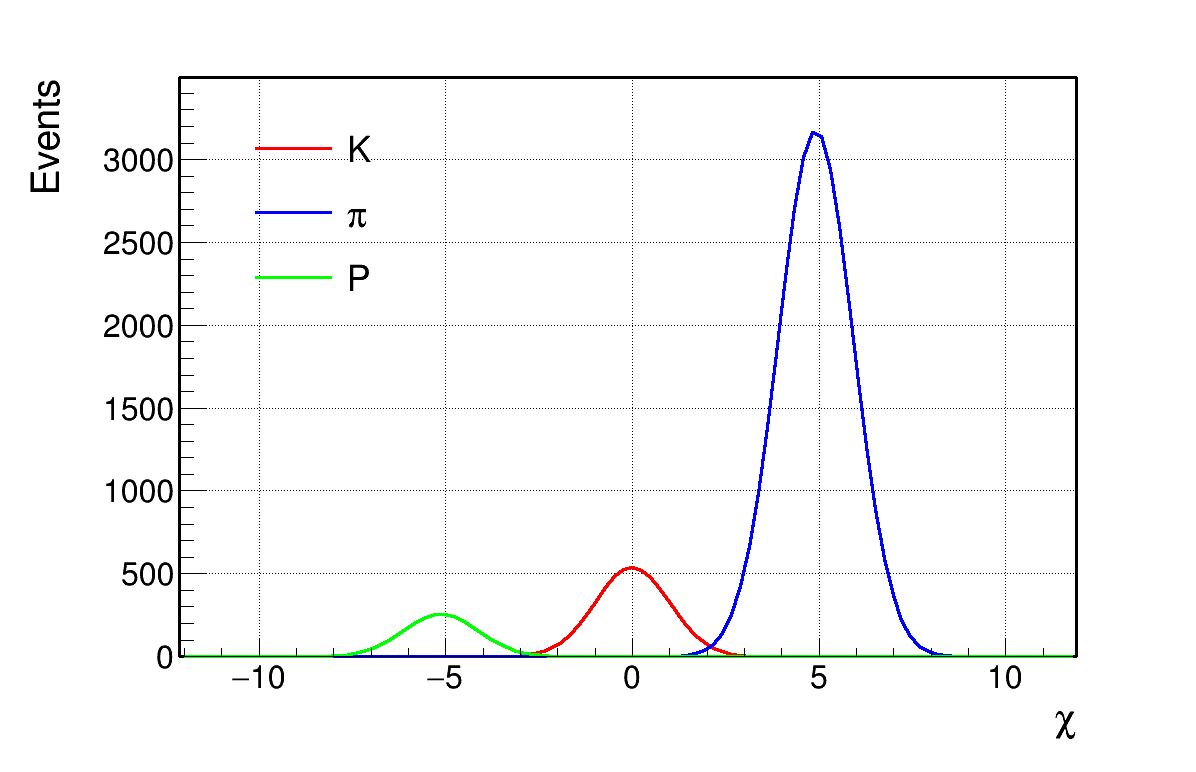}
    \caption{Ideal distributions of the three particle species for $p = 3~\mathrm{GeV}/c$ and $|\cos\theta| = 0.5$, where $\chi$ is obtained from the combination of the TPC, ITK, and OTK.}
    \label{fig:separation_ability}
\end{figure}

\begin{figure}[t]
    \centering
    \includegraphics[width=\linewidth]{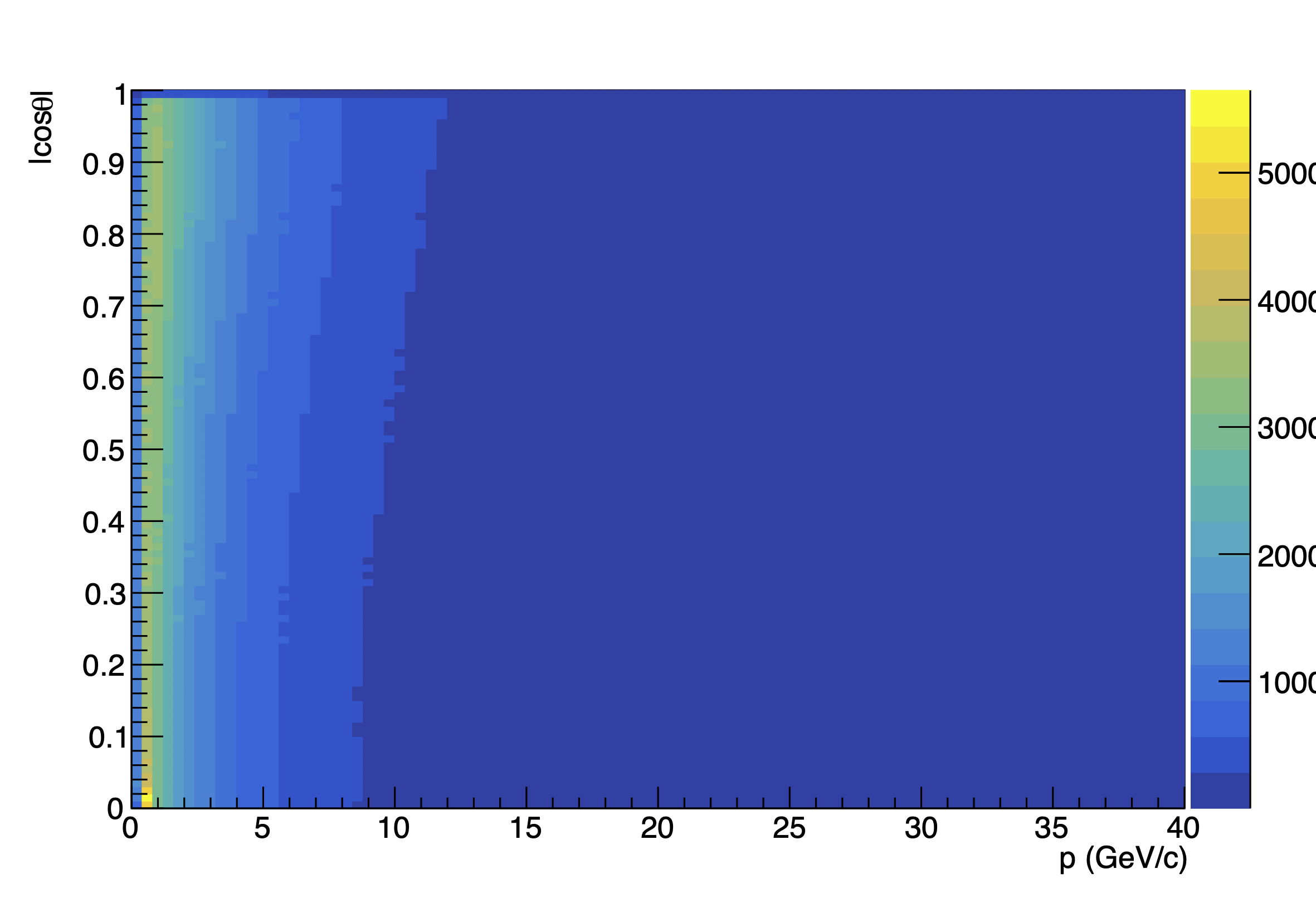}
    \caption{Kaon distribution map as a function of $p$ and $|\cos\theta|$.}
    \label{fig:mapK}
\end{figure}

The performance of PID is highly sensitive to the choice of regions in $\chi$ space used to assign particle types. For a representative choice of $p$ and $\cos\theta$, Fig.~\ref{fig:separation_ability} shows the distributions of the three particle species $K$, $\pi$, and $p$ as a function of $\chi$. They are modeled as Gaussian distributions with unit width. The separations between the Gaussian means are calculated from the separation power, and the relative amplitudes are obtained from the simulation results. A map of the kaon distribution as a function of $p$ and $\cos\theta$ is shown in Fig.~\ref{fig:mapK} as an example.

\begin{figure}
    \centering
    \includegraphics[width=\linewidth]{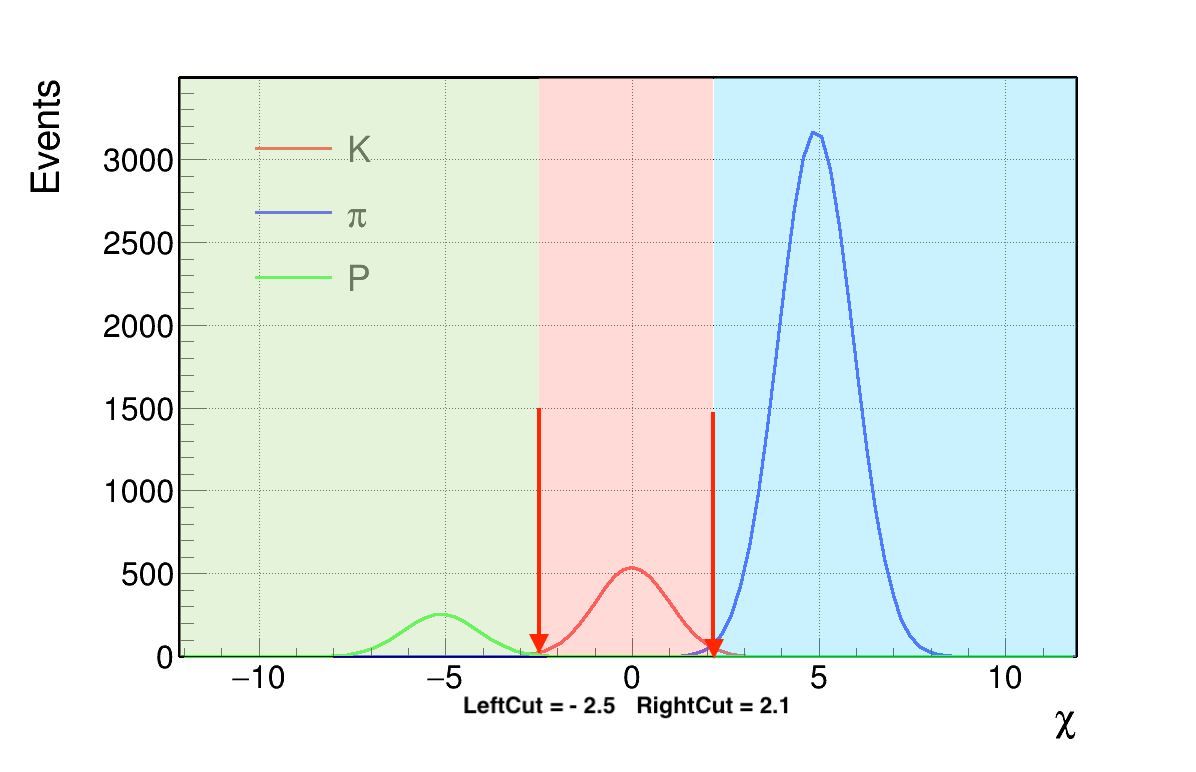}
    \caption{Illustration of the threshold scan based on Fig.~\ref{fig:separation_ability}.}
    \label{fig:efficiencytimespurity}
\end{figure}

In Fig.~\ref{fig:efficiencytimespurity}, the two red arrows mark the thresholds that partition the $\chi$ distribution into three regions. For every $(p, |\cos\theta|)$ bin, a track is classified as the particle type associated with the region in which its $\chi$ value falls, i.e. proton in the green region, kaon in the red region, and pion in the blue region. The threshold positions are optimized by scanning all possible values and selecting those that maximize the product of kaon identification efficiency and purity. This scan is performed for each $(p, |\cos\theta|)$ bin using hadronic $Z$-pole events, and the optimal thresholds are stored as a lookup table.

\section{Results}
\label{sec:results}

\begin{figure}[t]
    \includegraphics[width=\linewidth]{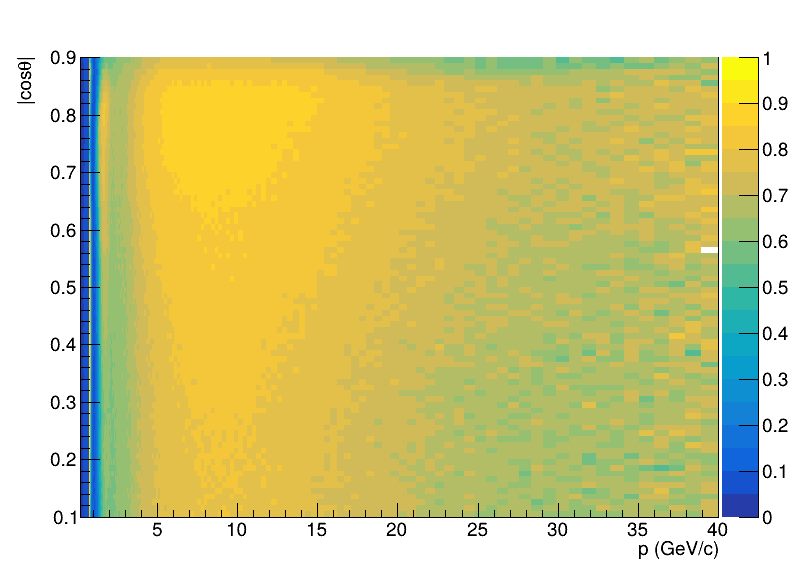}
    {\centering (a) \par}
    \vspace{1em}
    \includegraphics[width=\linewidth]{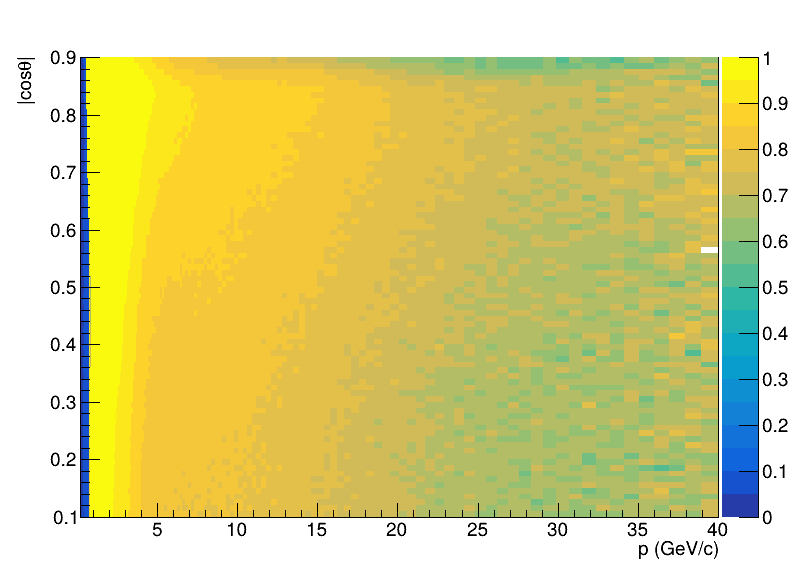}
    {\centering (b) \par}
    \vspace{1em}
    \includegraphics[width=\linewidth]{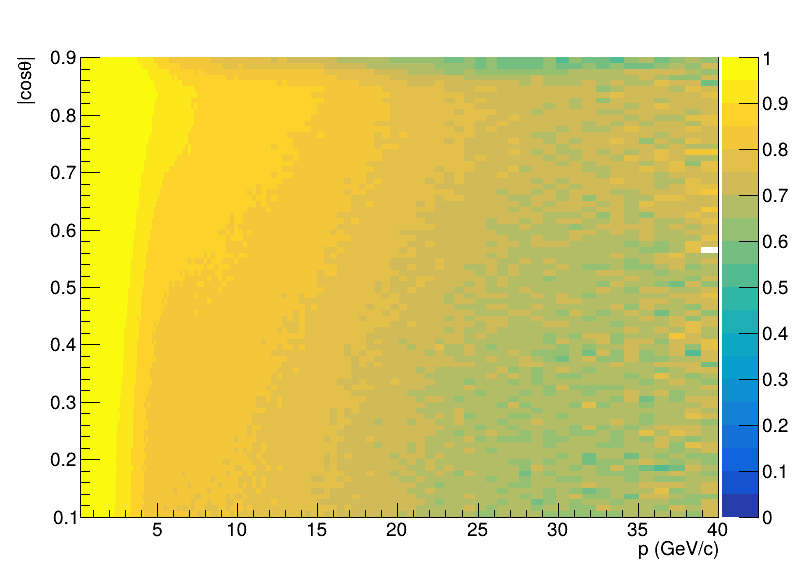}
    {\centering (c) \par} 
    \caption{Comparison of the product of kaon identification efficiency and purity.\\
    (a) The product with only data from TPC.\\
    (b) The product with combined data from TPC and OTK.\\
    (c) The product with combined data from ITK, TPC and OTK.}
    \label{fig:efficiencytimespurity_compare}
\end{figure}

As shown in panel (a) of Fig.~\ref{fig:efficiencytimespurity_compare}, the TPC alone exhibits limited PID performance for tracks with momenta below 3~$\mathrm{GeV}/c$. This limitation arises from increased ionization fluctuations and the overlap of dN/dx response distributions among different particle species at low momentum, thereby reducing separation power.

Panel (b) illustrates that the inclusion of the OTK significantly enhances PID performance in the intermediate momentum range without compromising performance at higher momenta. The OTK provides complementary ToF information that improves overall discrimination. However, due to its large radial location, particles with momenta below approximately 1~$\mathrm{GeV}/c$ typically fail to reach it, restricting the effective range of the TPC+OTK configuration to $p \gtrsim 1~\mathrm{GeV}/c$.

To extend PID coverage into the sub-GeV domain, the ITK is introduced, as shown in panel (c). The addition of the ITK enables PID of low-momentum tracks that are inaccessible to the OTK, thereby allowing continuous PID over the full momentum range when combined with TPC and OTK. Nevertheless, the performance gain at low momentum comes at a cost: as momentum increases, the ToF differences between particle species decrease, leading to larger relative timing uncertainties. As a result, the ITK contribution becomes marginal at high momentum because the ToF differences shrink, so the PID performance in this region is largely dominated by the TPC.

\begin{figure}[t]
    \centering
    \includegraphics[width=\linewidth]{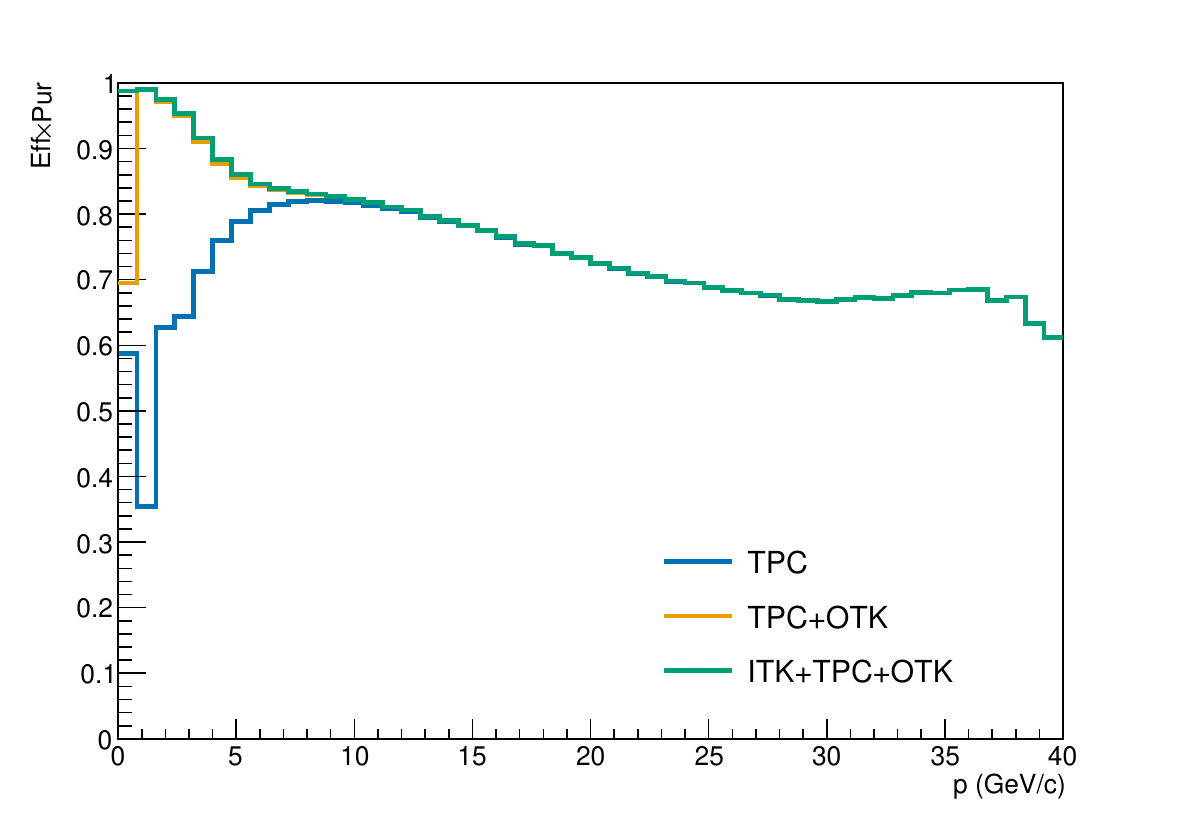}
    {\centering (a) \par}
    \vspace{1em}
    \includegraphics[width=\linewidth]{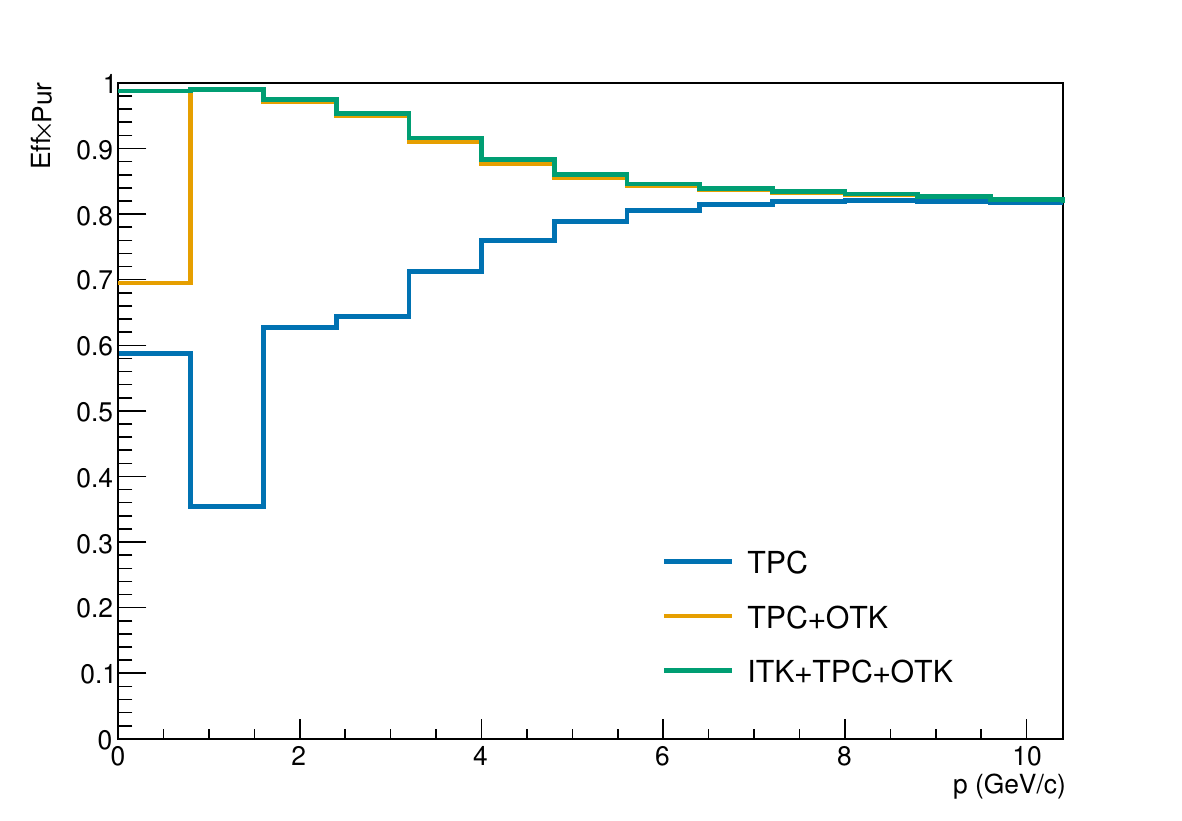}
    {\centering (b) \par}
    \vspace{1em}
    \caption{Product of kaon identification efficiency and purity as a function of momentum.\\
    (a) Full momentum range.\\
    (b) Zoomed view in the range 0–10~$\mathrm{GeV}/c$.}
    \label{fig:weieffpur}
\end{figure}

Fig.~\ref{fig:weieffpur} presents a direct comparison of the PID performance from the three detector combinations as a function of track momentum. The plot highlights their relative effectiveness across different momentum regions. The ITK+TPC+OTK configuration provides the best overall PID performance across the full momentum range: for $p < 1 ~\mathrm{GeV}/c$, the ITK is employed to ensure coverage of low-momentum tracks; in the intermediate region ($1 < p < 5~\mathrm{GeV}/c$), OTK yields superior performance; while for $p > 5~\mathrm{GeV}/c$, the TPC provides the dominant contribution.

Quantitatively, Table~\ref{tab:effpur} presents the kaon identification efficiency, purity, and their product for different detector configurations over the full momentum range. In the low-momentum region ($0.25 < p < 1.0~\mathrm{GeV}/c$), the TPC-only configuration suffers from a critically low purity of only 6.93\%, despite a reasonable efficiency of 91.6\%. The OTK cannot provide effective separation due to geometrical constraints; consequently, the efficiency–purity product of the TPC+OTK configuration is only 8.3\% in this low-momentum region. The ITK plays a crucial role in this regime. By combining the ITK with the OTK and TPC, the kaon efficiency reaches a high level of 99.9\%, while the purity achieves 81.9\%.

In the intermediate momentum range ($1.0 < p < 5.0~\mathrm{GeV}/c$), a significant performance leap is observed with the addition of OTK. While the TPC-only case maintains a moderate efficiency (88.3\%), its purity remains inadequate at 44.6\%. By incorporating OTK, the purity sharply increases to 98.0\%, yielding an efficiency–purity product of 96.3\%. In the presence of a high mis-identification rate caused by the overlapping response curves in the dN/dx observable, the TOF information provided by the OTK offers effective kaon separation.

At $5.0 < p < 40~\mathrm{GeV}/c$, the TPC emerges as the dominant contributor to the PID process. The performance becomes relatively stable across all configurations, with efficiencies consistently around 93\% and purity near 89\%. In this regime, the marginal gain from adding ITK or OTK is limited (approximately 2\% in purity). As the momentum increases, the ToF differences between particle species decrease, which reduces the timing-based separation capability of the ITK and OTK. In this momentum regime, the TPC provides the primary contribution to the identification.

Ultimately, while the TPC performs excellently within the $5\text{--}40\text{ GeV}/c$ range and the OTK stands out within $1\text{--}5\text{ GeV}/c$, the majority of kaons and pions are distributed at lower momenta. This results in an overall purity of only 23.2\% for the TPC across the $0.25\text{--}40\text{ GeV}/c$ range, which improves to just 30.5\% when combined with the OTK. However, the combination of ITK, OTK, and TPC significantly elevates the purity to 85.6\% within this range, while achieving an efficiency of 97.1\%, corresponding to an efficiency–purity product of 83.1\%.

\begin{table}[t]
    \centering
    \caption{Comparison of kaon identification efficiency, purity and their products for different combinations.
    }
    \label{tab:effpur}
    \vspace{1em}
    \centerline{\textbf{(a) $0.25 < p < 1.0~\mathrm{GeV}/c$}}
    
    \begin{tabular}{cccc}
        \toprule
        Category & efficiency & purity & eff*pur \\
        \midrule
        TPC        & 0.916      & 0.069 & 0.063  \\
        TPC+OTK        & 0.994      & 0.083 & 0.083  \\
        TPC+OTK+ITK        & 0.999      & 0.819 & 0.819  \\
        \bottomrule
    \end{tabular}
    
    \vspace{1em}
    \centerline{\textbf{(b) $1.0 < p < 5.0~\mathrm{GeV}/c$}}
    
    \begin{tabular}{cccc}
        \toprule
        Category     & efficiency & purity & eff*pur \\
        \midrule
        TPC          & 0.883      & 0.446 & 0.394  \\
        TPC+OTK      & 0.983      & 0.980 & 0.962  \\
        TPC+OTK+ITK  & 0.984      & 0.981 & 0.965  \\
        \bottomrule
    \end{tabular}
    
    \vspace{1em}
    \centerline{\textbf{(c) $5.0 < p < 40~\mathrm{GeV}/c$}}
    
    \begin{tabular}{cccc}
        \toprule
        Category     & efficiency & purity & eff*pur \\
        \midrule
        TPC          & 0.927      & 0.875 & 0.811  \\
        TPC+OTK      & 0.931      & 0.891 & 0.830  \\
        TPC+OTK+ITK  & 0.931      & 0.892 & 0.831  \\
        \bottomrule
    \end{tabular}
    
    \vspace{1em}
    \centerline{\textbf{(d) $0.25 < p < 40~\mathrm{GeV}/c$}}
    
    \begin{tabular}{cccc}
        \toprule
        Category     & efficiency & purity & eff*pur \\
        \midrule
        TPC          & 0.903      & 0.232 & 0.210  \\
        TPC+OTK      & 0.969      & 0.305 & 0.295  \\
        TPC+OTK+ITK  & 0.971      & 0.856 & 0.831  \\
        \bottomrule
    \end{tabular}
\end{table}

\section{Conclusion}
\label{sec:conclusion}

In this study, we performed a detailed evaluation of the PID performance at the CEPC using fully simulated hadronic $Z$-pole events. By combining dN/dx measurements from the TPC with ToF information from the ITK and OTK, we constructed a unified PID discriminant to distinguish among charged hadrons. Optimal PID regions in the discriminant space were determined by maximizing the product of kaon efficiency and purity across bins of momentum.

The results show that the TPC alone provides effective PID at high momentum but suffers from limited separation at low momentum because of increased ionization fluctuations and overlapping response curves in the dN/dx observable. The addition of OTK timing information significantly improves the performance in the intermediate-momentum region, while an upgraded ITK timing layer restores PID capability below 1~GeV/$c$. Over the full momentum range $0.25 < p < 40$~GeV/$c$, the combined ITK+TPC+OTK configuration achieves a kaon identification efficiency of 97.1\% and a purity of 85.6\%, corresponding to an efficiency--purity product of 83.1\%.

Future studies will investigate the impact of ITK and OTK timing resolutions on PID performance in greater detail, which will guide further detector design optimization and calibration strategies.

\section*{Acknowledgments}
We thank Xiaotian Ma and Chenguang Zhang for their help, and we gratefully acknowledge the contribution of Kaili Zhang, who has since passed away, in producing the samples. This work was supported by the Open Fund of the China Spallation Neutron Source Songshan Lake Science City (Grant No.~DG24313513), New Initiative Action B of CAS(Grant No.~292024000087), NSFC Basic Science Center Program (Grant No.~12188102), the Young Talents of National Talent Support Programs (Grant No.~24Z130300579).

\end{document}